\documentclass[aps,prl,reprint,longbibliography,superscriptaddress,floatfix]{revtex4-2}
\usepackage{graphicx}
\usepackage{bm}
\usepackage{amsmath}
\usepackage{amssymb}
\usepackage[utf8]{inputenc}
\usepackage[T1]{fontenc}
\usepackage[dvipsnames]{xcolor}
\usepackage{newunicodechar}
\newcommand{\dw}{{d_w}}
\newcommand{\df}{{d_f}}

\begin{document}
\title{Heterogeneous Mean First-Passage Time Scaling in Fractal Media}

\author{Hyun-Myung Chun}
\affiliation{School of Physics, Korea Institute for Advanced Study, Seoul 02455, Korea}
\author{Sungmin Hwang}
\affiliation{Capital Fund Management, 75007 Paris, France}
\author{Byungnam Kahng}
\affiliation{Center for Complex Systems Studies, and KENTECH Institute for Grid Modernization, Korea Institute of Energy Technology, Naju 58217, Korea}
\author{Heiko Rieger}
\affiliation{Center for Biophysics and Department of Theoretical Physics, Saarland University, 66123 Saarbr\"ucken, Germany}
\affiliation{Lebniz-Institute for New Materials INM, 66123 Saarbr\"ucken, Germany}
\author{Jae Dong Noh}
\affiliation{Department of Physics, University of Seoul, Seoul 02504, Korea}

\date{\today}

\begin{abstract}
The mean first passage time~(MFPT) of random walks is a key quantity characterizing dynamic processes  on disordered media. In a random fractal embedded in the Euclidean space, the MFPT is known to obey the power law scaling with the distance  between a source and a target site with a universal exponent. We find that the scaling law for the MFPT is not determined solely by the distance  between a source and a target but also by their locations.  The role of a site in the first passage processes is quantified by the  random walk centrality. It turns out that the site of highest random walk centrality, dubbed as a hub, intervenes in first passage processes. We show that the MFPT from a departure site to a target site is determined by a competition between direct paths and indirect paths detouring via the hub. Consequently, the MFPT displays a crossover scaling between a short distance regime, where direct paths are dominant, and a long distance regime, where indirect paths are dominant. The two regimes are characterized by power laws with different  scaling exponents.  The crossover scaling behavior is confirmed by extensive numerical calculations of the MFPTs on the critical percolation cluster in two dimensional square lattices. \end{abstract}

\maketitle
{\it Introduction} -- 
Random walks are fundamental for stochastic processes, such as transport, search, and spreading. While random walks on regular lattices have long been studied~\cite{Hughes.1995}, there has been an ever-increasing interest in the topic incorporating structural disorder of the underlying substrate~\cite{Havlin.2002}, geometric confinement~\cite{Benichou.2010}, stochastic resetting~\cite{Evans.2020}, non-Markovian dynamics~\cite{Barbier-Chebbah.2022}, and many more. 

An important quantity characterizing random walks (RWs) is the first passage time (FPT) distribution and the mean first passage time (MFPT)~\cite{Hughes.1995,Redner2001}. Scaling properties of the FPT and MFPT reflect the interplay between the RW-dynamics and geometric properties of the underlying substrate. For example, on infinite lattices, the FPT distribution follows a power law with a universal exponent~\cite{Hughes.1995,Redner2001}. Generally, in finite scale-invariant media, the MFPT $T(r)$ between two sites at a distance $r$ is known to obey the scaling law~\cite{Condamin.2007,reuveni.2010,Roberts.2011}
\begin{equation}
    T(r) \sim \left\{
        \begin{aligned}
            N r^{\dw -\df} &,\quad \mbox{ for } \dw > \df \\
            N \ln r &,\quad \mbox{ for } \dw = \df \\ 
            N &,\quad \mbox{ for } \dw < \df 
    \end{aligned} \right. ,
    \label{T_Condamin}
\end{equation}
where $N$ is the total number of sites, $\df$ is the fractal dimension of the medium, and $d_w$ is its walk dimension. It is remarkable that the scaling law is governed by only one universal exponent, $\theta=\dw-\df$.
On the other hand, on a highly heterogeneous graph, the MFPT displays a more complex scaling behavior~\cite{Tejedor.2009,Agliari.2010,Tejedor.2011,Hwang.201236k}. 
In a scale-free network characterized by a power-law distribution of local connectivity of each site, the FPT and the MFPT averaged over source sites display a target site dependent scaling behavior~\cite{Hwang.201236k}.
Generally, in heterogeneous media, the MFPT from site $i$ to $j$ could be very different from the MFPT from $j$ to $i$: for undirected graphs, one can assign a potential-like quantity, called the RW centrality~(RWC), to each site~\cite{Noh.2004}. Since the MFPT between two sites in either direction differs by the difference in their inverse RWCs (see below), a wide distribution of the RWCs could lead to a source-target specific, or heterogeneous, scaling of the MFPT, which is what we will address in this paper.

To this purpose, we reconsider the scaling law in Eq.~\eqref{T_Condamin} for two-dimensional~(2D) critical bond percolation clusters. We will show that despite a homogeneous local connectivity distribution, the MFPT displays a heterogeneous scaling behavior characterized by a site-dependent scaling exponent and an intriguing crossover scaling, for which the site with the highest RWC is responsible. RWs on critical percolation clusters have long been studied~\cite{d'Auriac.1983, Alexander.1982, Havlin.2002, benichou2008zero, condamin2008probing}, but a site-dependent or heterogeneous scaling has not been reported yet. Our work also sheds light on the role of the RWC for RWs in disordered media. 

{\it Random walk centrality} --
We consider an undirected graph consisting of $N$ sites, whose connectivity is  represented with a symmetric adjacency matrix $\mathsf{A} = \mathsf{A}^T$ whose matrix elements $A_{ij}$ are $0$ or $1$ indicating the absence or presence of an edge between sites $i$ and $j$~\footnote{The formalism in the paper is also valid in a weighted graph with non-binary adjacency matrix elements.}, respectively. The number of edges attached to a site $i$ is its degree and is given by $k_i = \sum_{j}A_{ij}$. A discrete time RW on the graph  is defined by the transition matrix $\mathsf{W} = \mathsf{K}^{-1} \mathsf{A}$, where  $\mathsf{K}$ is a diagonal matrix with $K_{ij} = \delta_{ij} k_i$. That is, a random walker at site $i$ jumps to site $j$ with the probability $W_{ij} = A_{ij}/{k_i}$ in a unit time step $\Delta t = 1$. The transition matrix has the left row eigenvector $\langle \bm{\pi}| = (\pi_1, \cdots, \pi_i,\cdots, \pi_N)$ with $\pi_i = k_i / (\sum_j k_j)$ and the right column eigenvector $|\bm{1}\rangle  =  (1,1,\cdots, 1)^T$, both with eigenvalue $\lambda=1$. The left eigenvector corresponds to the steady state probability  distribution~\cite{Noh.2004}.

A general theoretical framework for studying discrete time random walks  has been formulated some time ago~\cite{Kemeny.1976,Hughes.1995,Noh.2004}. There the MFPT from site $i$ to $j$ is given by~\cite{Noh.2004} \begin{equation}
    T_{ij} = \frac{R_{jj} - R_{ij}+\delta_{ij}}{\pi_j}  ,
    \label{MFPT_formal}
\end{equation}
where the matrix $\mathsf{R}$ is called the group generalized inverse of $(\mathsf{I-W})$~\cite{Jr..1975,Hunter.2014} and given by  $\mathsf{R} \equiv \sum_{t=0}^\infty ( \mathsf{W}^t - |\bm{1}\rangle  \langle \bm{\pi}|)$~\footnote{The matrix $(\mathsf{I-W})$ is not invertible because one of the eigenvalues of $\mathsf{W}$ equals to 1. The group generalized inverse is defined as $\sum_{n=2}^N \frac{1}{1-\lambda_n} |\lambda_n\rangle\langle \lambda_n|$, where $\lambda_n$, $|\lambda_n\rangle$, $\langle \lambda_n|$ are the $n$-th eigenvalue, right eigenvector, and left eigenvector of $\mathsf{W}$, respectively. The sum excludes the eigenstate with $\lambda_1 = 1$.}.   Condamin {\it et al.}~\cite{Condamin.2007} noticed that $R_{ij}$ is dominated by the term $\sum_{t=0}^\infty W(j,t|i)$ where $W(j,t|i) \equiv  (\mathsf{W}^t)_{ij}$ is the probability to find the walker at site $j$ in $t$ steps when it started at site $i$. Assuming the scaling form $W(j,t|i) = t^{-\df/\dw} \Pi\left(\frac{r_{ij}}{t^{1/\dw}}\right)$ with $r_{ij}$ being the Euclidean distance between $i$ and $j$~\cite{Havlin.2002}, they derived the scaling law in Eq.~\eqref{T_Condamin}~\cite{Condamin.2007}.

The formal expression in Eq.~\eqref{MFPT_formal} has a deeper implication when the transition probabilities satisfy  the detailed balance condition, $\pi_i W_{ij} = \pi_j W_{ji}$ for all $i$ and $j$, which holds for RWs on undirected graphs.  Then, one can assign an RWC $C_i \equiv  \pi_i/R_{ii}$ to each site $i$, relating the MFPTs $T_{ij}$ and $T_{ji}$ by~\cite{Noh.2004}
\begin{equation}
    T_{ij} - T_{ji} = C_j^{-1} - C_i^{-1} .
    \label{TijTji}
\end{equation}
The RWC is an indicator of the attractiveness of a site in the random walk process: a first passage to a higher RWC site from a lower RWC site takes less time than the first passage in the opposite direction.  The RWC has also been used to identify influential nodes in complex networks~\cite{Loecher.2007,Blochl.2011,Johnson.2019,oldham.2019,Riascos.2020}.  The inverse of the RWC $\alpha_i \equiv 1/C_i = R_{ii}/\pi_i$ is equal to the average MFPT to site $i$ from a random departure site $j$ sampled with  the steady state probability distribution,  $\alpha_i = \sum_{j\neq i}\pi_j T_{ji}$. It is also called the global mean first passage time~\cite{Tejedor.2009,zhang.2010}, or the accessibility index~\cite{Kirkland.2016}. Similarly, Kemeny's constant is defined as $K_i = \sum_{j\neq i} T_{ij} \pi_j = \sum_{j} R_{jj}$, which is independent of $i$ and a characteristic of an underlying graph~\cite{Kemeny.1976, Kirkland.2016,yilmaz2020kemeny}. 

{\it MFPT from hub and marginal site} -- 
In a disordered medium site-to-site fluctuations of the RWC may affect the scaling of the MFPT with the distance. We address this issue for the RW on the critical bond percolation cluster  of 2D square lattices~\cite{d'Auriac.1983,Alexander.1982,Havlin.2002,benichou2008zero,Benichou.2010}, which we generate by occupying bonds of a 2D $L\times L$ square lattice with periodic boundary conditions with the critical  occupation probability $p_c = 1/2$ and identifying the largest cluster. The 2D critical percolation cluster is a random fractal with the fractal dimension $\df = 91/48$~\cite{Stauffer.1992}. The walk dimension is known to be $\dw \simeq 2.87 > \df$~\cite{Majid.1984}.

It is computationally demanding to evaluate the RWC and the MFPT  for it requires to find the group generalized inverse of $\mathsf{I-W}$~\cite{Johnson.2019}.  We will adapt the numerical algorithm developed in Ref.~\cite{Hwang.2014}, which turns out to be extremely efficient.  It takes only a few minutes in an ordinary desktop computer to compute the RWC distribution for the critical percolation cluster of lattices of size $1024\times 1024$.  All numerical data for 2D percolation clusters  are obtained on a 2D lattice with $L=1024$, if not stated otherwise, and averaged over at least 2000 independent realizations of the critical percolation cluster.

\begin{figure}
\includegraphics*[width=\columnwidth]{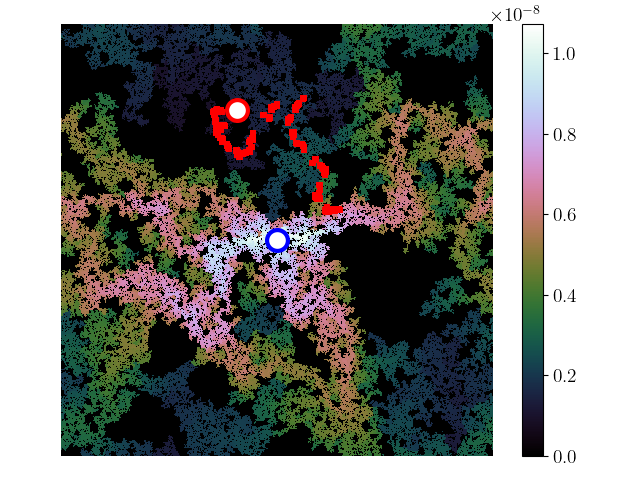}
\caption{RWC configuration on a 2D critical percolating cluster in a lattice of size $1024\times 1024$. The highest and lowest RWC sites are marked with blue~({\color{blue}$\bigcirc$}) and red~({\color{red}$\bigcirc$}) circles, respectively. Red line segments denote bridges between them~(see the main text). The black area represents sites that do not belong to the percolating cluster.}
\label{fig1}
\end{figure}

Fig.~\ref{fig1} illustrates an RWC configuration on a critical bond
percolation cluster. The RWC distribution is highly heterogeneous~(see
App.~A).  High RWC sites are clustered and spread out in a filamentous pattern, which is analogous to the backbone structure~\cite{Havlin.2002}.  This heterogeneity raises questions about the simple scaling of the MFPT  with a single scaling exponent as in Eq.~\eqref{T_Condamin}.

\begin{figure}
\includegraphics*[width=\columnwidth]{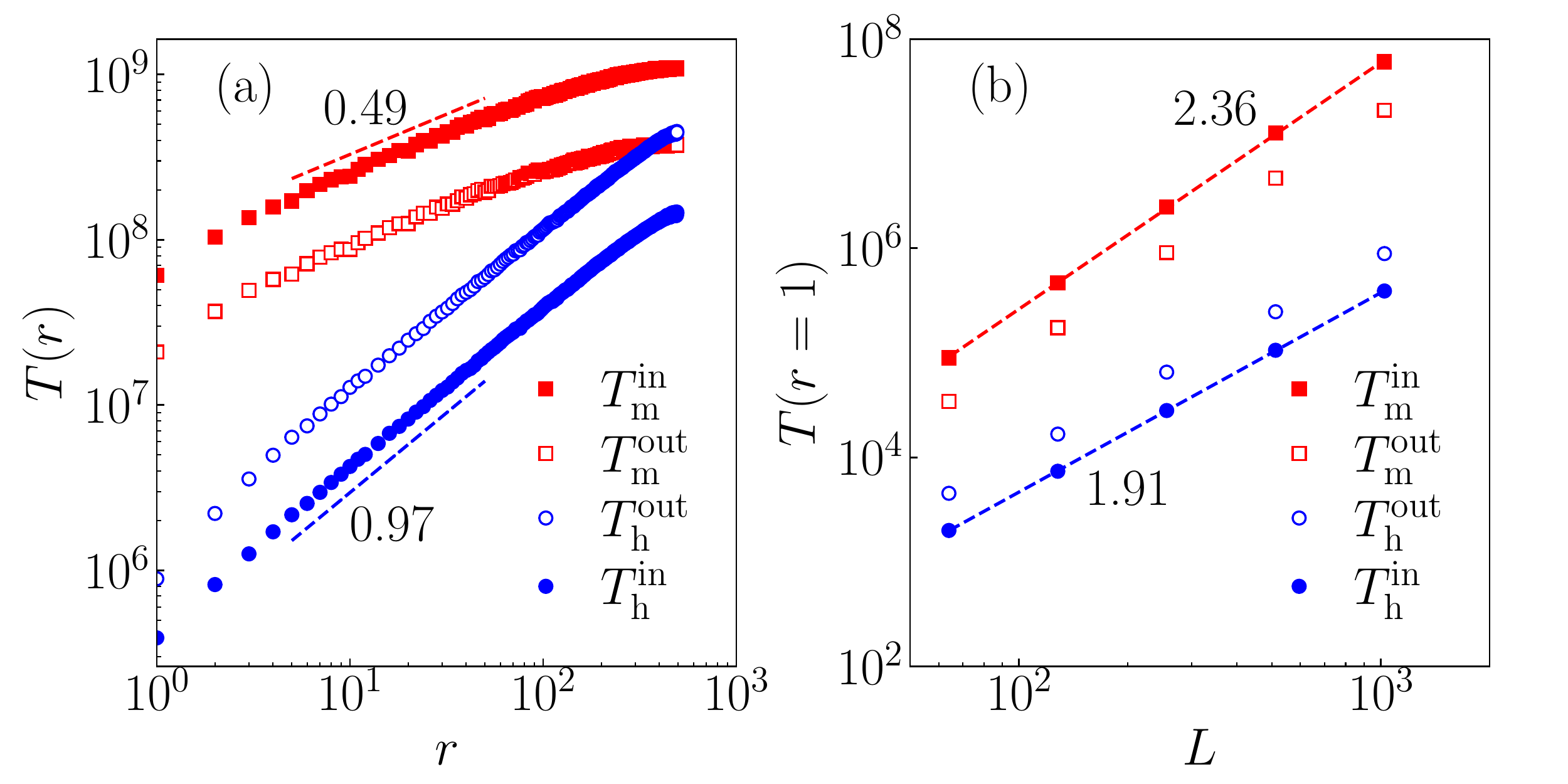}
\caption{(a) Average MFPT between the hub~(circular symbols) or the marginal site~(rectangular symbols) with the other sites. The outbound and inbound MFPTs are marked with empty and filled symbols, respectively. In (b), we plot the MFPTs from/to sites at a unit distance $r=1$ as a function of the linear system size  $64 \leq L \leq 1024$. The scaling exponents are obtained by fitting the data within the range indicated by the dashed lines.}
\label{fig2}
\end{figure}

To highlight the site dependence of the MFPT, we identify the {\em hub}~(highest RWC site) and the {\em marginal site}~(lowest RWC site),  and measure the average outbound/inbound MFPTs to/from  all the other sites at a given distance $r$. As shown in Fig.~\ref{fig2}, the outbound MFPT is larger than the inbound MFPT for the hub, and vice versa for the marginal site. The difference is exactly given by the difference in the inverse RWCs~(see Eq.~\eqref{TijTji}). The MFPTs scale algebraically with $L$ and $r$ as
\begin{equation}
    T \sim L^\Delta r^\theta 
    \label{T_Lr}
\end{equation}
with the scaling exponents $\Delta$ and $\theta$. Surprisingly, the hub and the marginal site are characterized by different scaling exponents 
\begin{equation}
    \theta = \left\{ 
        \begin{aligned} \theta_h \simeq 0.97(5) & \mbox{,\ for the hub}\\ 
    \theta_m \simeq 0.49(5) & \mbox{,\ for the marginal site.}  \end{aligned} \right.
    \label{theta_exp}
\end{equation}
The exponent $\theta_h$ associated with the hub is close to the exponent
$\dw - \df = 0.97(4)$ of Eq.~\eqref{T_Condamin} in value.  On the other
hand, the MFPT at the marginal site grows with a considerably smaller
exponent $\theta_m\simeq 0.49$~(see App.~B for detailed analysis).  The finite size scaling exponent $\Delta$ also varies. It takes on $\Delta_{h} \simeq 1.90$ for the hub, which is in agreement with the fractal dimension $\df=91/48$ of Eq.~\eqref{T_Condamin}~\cite{Condamin.2007}. The marginal site displays a stronger finite size effect with $\Delta_{m} \simeq 2.36 > \Delta_{h}$. 

We note that $\Delta_{h} + \theta_h \simeq  \Delta_{m} + \theta_m$. It is understood from the site independence of Kemeny's constant~\cite{Kirkland.2016}. Kemeny's constant evaluated at a site $i$ is given by $K_i = \sum_{j\neq i}T_{ij} \pi_j$. Since $\pi_j = a_j/N$ with $O(1)$ constant $a_j$, Kemeny's constant is approximated as the arithmetic average of outbound MFPTs to all the other sites. The scaling form in Eq.~\eqref{T_Lr} leads to $K \sim L^{\Delta+\theta}$. Thus, $\Delta + \theta$ should be the same at the sites obeying Eq.~\eqref{T_Lr}.  We also  note that the inbound and outbound MFPTs differ by a constant factor. From now on, we focus our study on the outbound MFPT. 

\begin{figure}
\includegraphics*[width=\columnwidth]{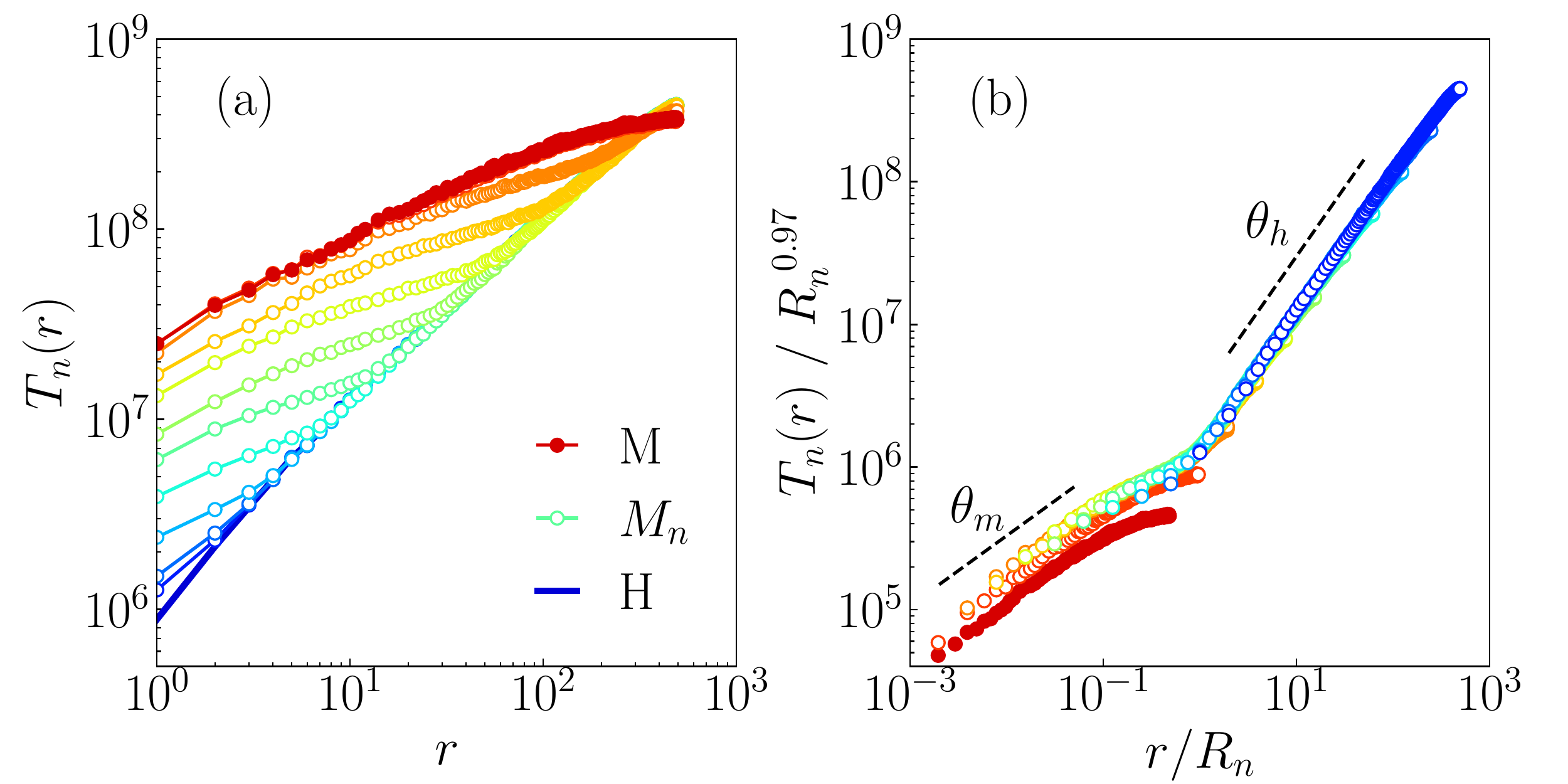}
\caption{(a) Outbound MFPT $T_n$ from the site $M_{n}$~(open symbols). Also shown is the MFPT from the hub~(H, thick line) and the marginal site~(M, filled symbols).  (b) Scaling plot of $T_n(r) / R_n^{\theta_h}$ vs $r / R_n$ with  $R_n = 2^{n-1}$. The dashed lines of slope $\theta_h$ and $\theta_m$ given in Eq.~\eqref{theta_exp} are guides to the eye.}
\label{fig3}
\end{figure}

{\em Crossover scaling of MFPT} -- 
The site-dependent scaling behavior is not limited to an exceptional outlier site: We consider the outbound MFPTs from a set of source sites $\{M_1, M_2, \cdots\}$ selected  hierarchically as follows.  We select the local minimum RWC site  $M_{n}$ among all sites within a circle of radius $R_n = 2^{n-1}$  centered at the hub. The outbound MFPT $T_n(r)$  from $M_n$ as a function of the distance $r$ to target sites is shown in  Fig.~\ref{fig3}(a). We find an interesting crossover of $T_n(r)$. It grows algebraically with $r$ with the exponent $\theta_m$ for $r\ll R_n$ and with the exponent $\theta_h$ for $r \gg R_n$. The crossover scaling behavior is summarized by the scaling form
\begin{equation}
    T_n(r) = N R_n^{\theta_h} \mathcal{F}(r / R_n),
    \label{crossover_scaling}
\end{equation}
where the scaling function $\mathcal{F}(x)$ behaves as  $\mathcal{F}(x \ll
1) \sim x^{\theta_m}$ and $\mathcal{F}(x\gg 1) \sim x^{\theta_h}$. The
crossover scaling is confirmed by the scaling plot shown in
Fig.~\ref{fig3}(b). The crossover scaling persists when one chooses a
source site at random among the sites at given distance from the hub~(see
App.~C).

Our numerical results highlight the role of the highest RWC site in the
random walk dynamics. Imagine an ensemble of the first passage events from a
source site $s$ to a target site $t$ at a distance $r_{s-t}$. Let $r_{s-h}$
be the distance from $s$ to the hub. When the target is farther than the
hub~($r_{s-t} \gg r_{h-s}$),  the ensemble is dominated by the paths
detouring via the hub. Consequently the MFPT follows the scaling  $T\sim
r^{\theta_h}$ with the scaling exponent  $\theta_h = \dw-\df$ irrespective
of $s$. On the other hand, when the target is closer than the hub~($r_{s-t}
\ll r_{h-s}$), the ensemble is dominated by direct paths and the MFPT
scaling law depends on the choice of $s$~{(see App.~D)}. The crossover may be overlooked when one measures the MFPT averaged over all pairs of source and target sites at a given distance.

One can understand the origin (and potential complications) of the  scaling law for $T(r)$ in Eq.~\eqref{T_Condamin}~\cite{Condamin.2007} by the following consideration: Given a source-target pair at a distance $r$, one partitions the entire graph into blocks of linear size $\xi_r \sim  r$, putting the source and target into the same block denoted as starting block. Each block has $N_r \sim r^{\df}$ sites and the total number of blocks is $\mathcal{N}_r \sim N / N_r \sim N r^{-\df}$.  If all blocks were statistically equivalent,  the RW would always spend $\tau_r \sim r^{\dw}$ time steps in a single block until it hops to a neighboring block.  With Eq.~\eqref{MFPT_formal} the return time  to the starting block is $T_{\rm ret.} \sim \tau_r \cdot (1/\pi_b)$,  with the  probability to be in one block $\pi_b\sim 1/\mathcal{N}_r \sim r^\df/N$, thus  $T_{\rm ret.} \sim N r^{\dw-\df}$.  The MFPT can then be estimated as  $T_{\rm ret.} / P_s$, with $P_s$ the probability to find the target site before leaving the starting block, which is $P_s = O(1)$ for $\dw > \df$  and $P_s \sim \tau_r / N_r \sim r^{\dw-\df}$ for $\dw < \df$.  This argument reproduces the scaling law  Eq.~\eqref{T_Condamin}, except for the marginal case $\dw=\df$, It clearly reveals that the simple scaling, $T(r)\sim r^\theta$ with a unique scaling exponent $\theta$, is based on the assumption  that the entire fractal lattice can be partitioned into homogeneous blocks.  Similar arguments may also lead to scaling laws for the higher moments of the FPT distribution, c.f. Ref.~\cite{Levernier.2018}. Our results presented in Fig.~\ref{fig3}, however, indicate that blocks are heterogeneous on all length scales.

This heterogeneity is further evidenced by the scaling behavior of the chemical distance (number of edges in the shortest path) $l(r)$ with respect to the Euclidean distance between two sites. The average chemical distance is known to scale as $l(r) \sim r^{d_{\rm min.}}$, with $d_{\rm min.} \simeq 1.14$ for the 2D critical percolation cluster (Sec.~6.6 of \cite{ben2000diffusion}). We discriminate again between the hub (h) and the marginal site (m) as starting site and found
\begin{equation}
l(r) \sim \left\{ 
        \begin{array}{llll}
        L^{\delta_h} r^{d_{\rm min.},h}, & \delta_{h} \simeq 0.0,
        & d_{{\rm min.},h} \simeq 1.11, \\ 
        L^{\delta_m} r^{d_{\rm min.},m}, & \delta_{m}\simeq 0.52,
        & d_{{\rm min.},m} \simeq 0.58.\\ 
     \end{array} \right.
    \label{lr_hm}
\end{equation}
The chemical distance $l_n(r)$ from the local minimum RWC site $M_n$  shows
again a crossover $l_n(r) = L^{\delta_h} R_n^{d_{{\rm min.}, h}}
\mathcal{G}(l/ R_n)$ for $1 \ll R_n \ll L$. We also looked at the MFPT as a
function of the chemical distance and observed a similar crossover
behavior~(see App.~E for the detailed analysis for the chemical distance scaling).

{\it Origin of crossover scaling} -- 
A fractal may comprise a subset of sites or bonds which is itself a fractal. For instance, 2D percolation clusters contain red bonds~(or cutting bonds) which themselves form a fractal with fractal dimension $d_{\rm red} = 3/4$~\cite{Coniglio.1999}. More generally, a percolation cluster has a meso-scale structure consisting of a backbone, red bonds, and dangling ends~\cite{Herrmann.1999, Havlin.2002}. This structural heterogeneity is the origin of the site-dependent scaling and the crossover scaling.

To support this claim, we study the distribution of {\em
bridges}~\cite{Tarjan.1974} between the hub and the marginal site in 2D
percolation clusters. A bond is defined to be a bridge if the marginal site
would be disconnected from the hub without it, represented by red lines in
Fig.1~\footnote{The bridges correspond to red bonds between the hub and the
marginal site. Imagine that all bonds are identical resistors and a voltage
drop is applied between the hub and the marginal site. The bridge bonds
carry all current in such an electric circuit, and the current stops flowing
when any of them are removed.}. We find that the total number of bridges
obeys the power law scaling $N_{\rm b} \sim L^{0.75}$~{(see
App.~F)}, which indicates that the marginal site is located deep within a dangling end. Moreover, as Fig.~\ref{fig1} illustrates, bridges are predominantly distributed near the marginal site. Consequently, random walks from the marginal site are {\em quasi-one-dimensional} along a path consisting mainly of the bridges. The MFPT from a marginal site to another separated by a chemical distance $l$ then scales like the 1D random walk MFPT $T(l) \sim N l^1$. Since the chemical distance $l$ scales with the Euclidean distance $r$ as given in Eq.~\eqref{lr_hm}, we obtain
\begin{equation}\label{conjecture}
    T(r) \sim L^{d_f + \delta_m} r^{d_{{\rm min.},m}} .
\end{equation}
Note that $d_f+\delta_m \simeq 2.42$ and $d_{{\rm min.},m}\simeq 0.58$, which are close to $\Delta_m \simeq 2.36$ and $\theta_m \simeq 0.49$, respectively (c.f. Eq.~\eqref{theta_exp}). Our argument reveals that the quasi-one-dimensional structure of bridges near the marginal site is responsible for the crossover scaling. It also predicts, at least approximately, the scaling exponents $\Delta_m$ and $\theta_m$ in terms of the geometric quantities $\delta_m$ and $d_{{\rm min.},m}$.

\begin{figure}
    \includegraphics*[width=\columnwidth]{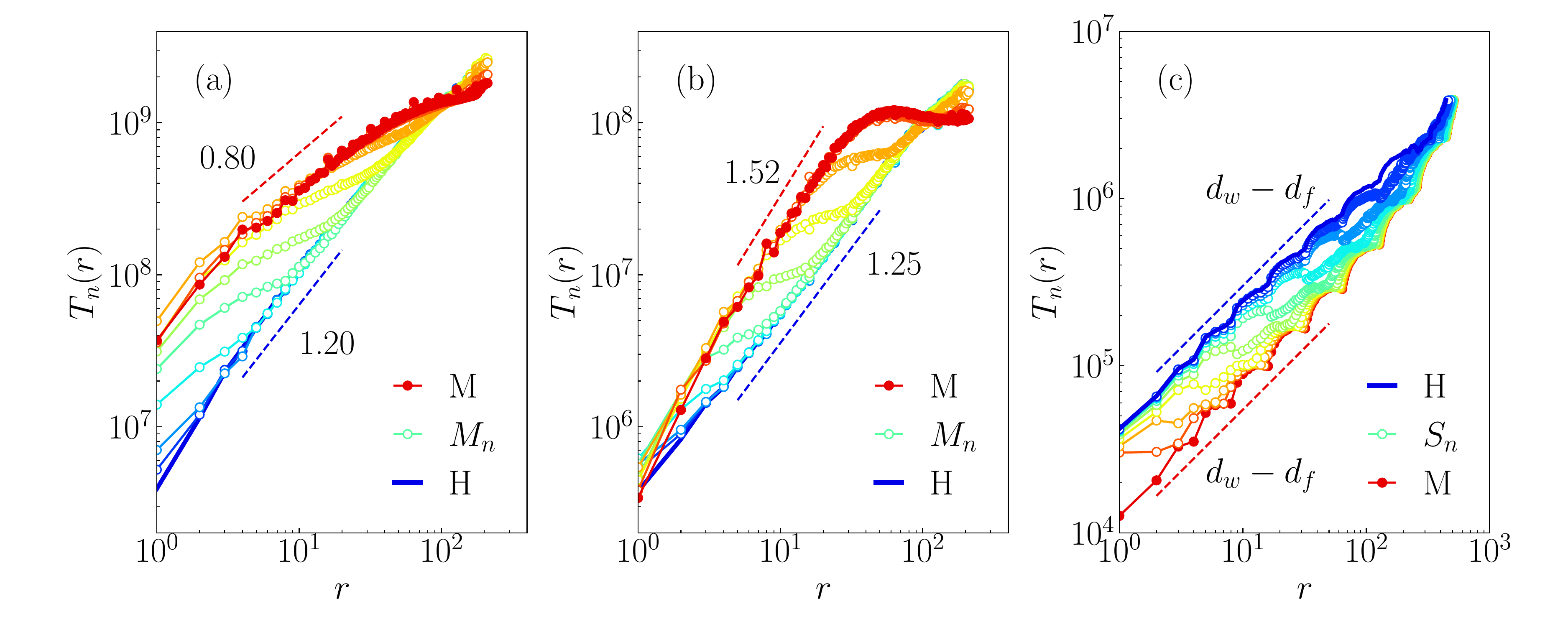}
    \caption{Crossover scaling of the MFPT for (a) 3D critical percolation clusters, (b) 3D random walk trails, and (c) the Sierpi\'nski gasket.}
    \label{fig4}
\end{figure}

{We performed similar studies for 3D percolation
clusters~\cite{Lorenz.1998}, random walk trails~\cite{Falconer.1985}, and the
Sierpi\'nski gasket. We observe the described crossover scaling for the
random fractals as shown in Fig.~\ref{fig4}. 
We point out that for the deterministic fractal shown here~(Sierpi\'nski
gasket), the scaling exponents are site-independent~(see App.~G for further discussion).}

{\it Conclusion} -- We report in this paper, for the first time, heterogeneous scaling behavior of the MFPT of RWs on a random fractal, the critical percolation cluster in 2D. MFPTs measured  from the hub and measured from the marginal site as a function of the distance of the target site obey power law scaling with distinctively different exponents and the distance dependence of MFPTs from general starting sites  shows a striking crossover. 

Heterogeneous behavior of various observables in disordered systems is expected, as for instance dynamical heterogeneities in glassy systems~\cite{Kob.1997,*Kegel.2000} or in the context of Griffiths singularities in  strongly disordered systems~\cite{Igloi.2005} like the transverse Ising chain~\cite{Fisher.1992,*Fisher.1995,*Igloi.1998a} or the Sinai walk~\cite{Fisher.1998,*Igloi.1998b}. A lack of self-averaging is a prominent consequence  of this spatial heterogeneity~\cite{Aharony.1996,*Wiseman.1998}, and it manifests itself in the quantities we looked at. But different power laws for different  regions in the system, as we find them for regions close to the hub and close to the marginal site, have, to our knowledge, not been reported before. An important consequence of our results is that scaling theories for the MFPT that are based on an explicit or hidden spatial homogeneity assumption  should be considered more carefully. 

{
The origin of the strong heterogeneity in the MFPT can be traced
back to the broad distribution of the RWC. Our results could be generalized
to, and are relevant for, a larger class of heterogeneous media embedded
in real space, like diffusion-limited aggregation, random resistor networks,
lattice animals, and so on. We also speculate that heterogeneous scaling
could occur in multifractal systems characterized by a continuous spectrum of 
fractal dimension. Our results also suggest that the RWC distribution is
important to understand information spreading dynamics on complex networks. 
}
 
\begin{acknowledgments}
J.D.N was supported by the National Research Foundation of Korea (NRF) grant
funded by the Korea government (MSIP) (Grants No. 2019R1A2C1009628). HR
acknowledges financial support  by the German Research Foundation (DFG)
within the Collaborative Research Center SFB 1027-A3 and INST 256/539-1. 
H.-M.C. was supported by
a KIAS Individual Grant (PG089401) at Korea Institute for Advanced Study.
B.K was supported by the National Research Foundation of Korea by Grand No.
NRF-2014R1A3A2-069005 and RS-2023-00279802 and the KENTECH Research Grant
No. KRG-2021-01-007.
\end{acknowledgments}
\bibliography{rwc2}

\renewcommand{\theequation}{S\arabic{equation}}
\renewcommand{\thefigure}{S\arabic{figure}}
\setcounter{equation}{0}
\setcounter{figure}{0}
\setcounter{secnumdepth}{2}
\onecolumngrid
\newpage
\begin{center}
{\bf\large Supplemental Materials}\\
\vspace{4mm}
\setcounter{page}{1}

Hyun-Myung Chun${}^{1}$, Sungmin Hwang${}^{2}$, Byungnam Kahng${}^{3}$,
Heiko Rieger${}^{4, 5}$, and Jae Dong Noh${}^{6}$\\
\vspace{2mm}

${}^{1}${\it School of Physics, Korea Institute for Advanced Study, Seoul 02455, Korea}\\

${}^{2}${\it Capital Fund Management, 75007 Paris, France}\\

${}^{3}${\it Center for Complex Systems Studies, and KENTECH Institute for Grid Modernization, Korea Institute of Energy Technology, Naju 58217, Korea} \\

${}^{4}${\it Center for Biophysics and Department of Theoretical Physics, Saarland University, 66123 Saarbr\"ucken, Germany}\\

${}^{5}${\it Lebniz-Institute for New Materials INM, 66123 Saarbr\"ucken, Germany}\\

${}^{6}${\it Department of Physics, University of Seoul, Seoul 02504, Korea}\\
\end{center}

\twocolumngrid

\appendix
\section{Distribution of the Random Walk Centrality}\label{App:RWCdistribution}
\begin{figure}[b]
    \includegraphics*[width=\columnwidth]{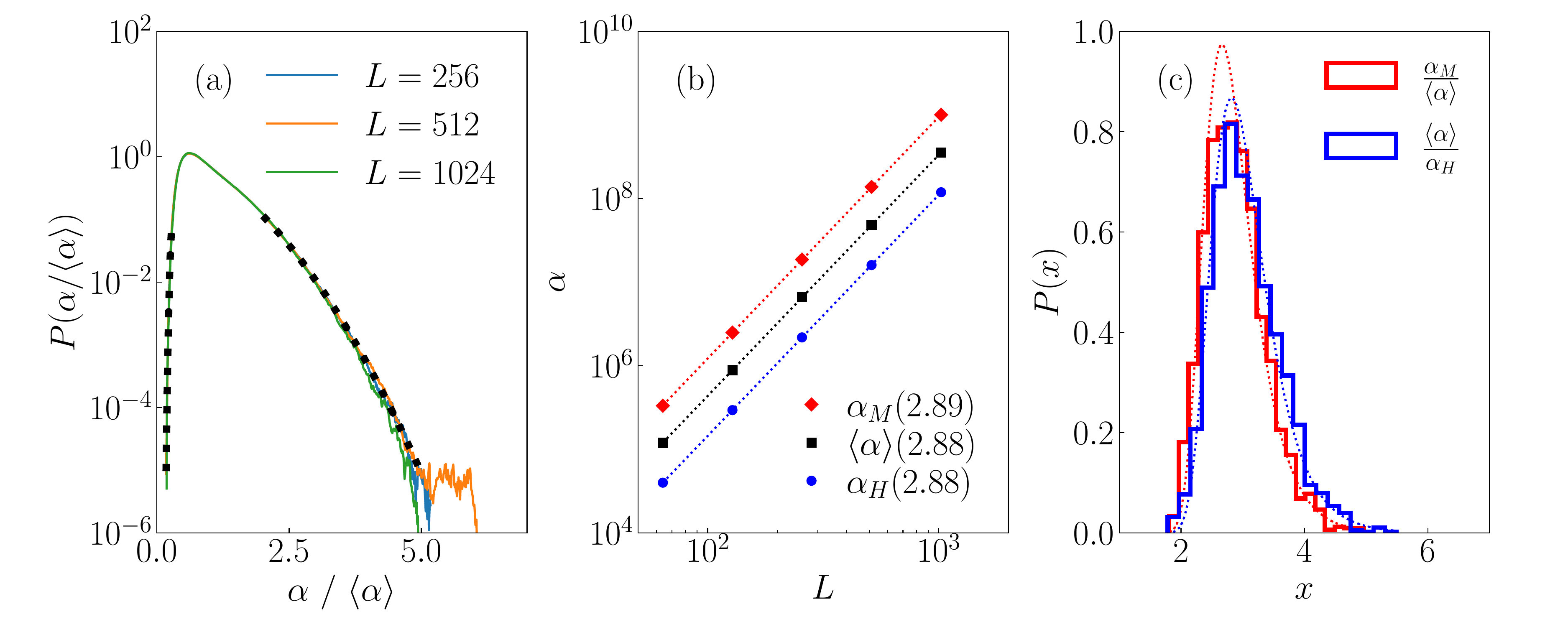}
    \caption{(a)~Distribution functions of the normalized accessibility index. The tails are fitted well by Gaussian functions~(dashed lines) with different stiffness on either side. (b)~System size dependence of the accessibility indices of the hub~($\alpha_H$),  the marginal site~($\alpha_M$), and the mean  value~($\langle\alpha\rangle$). The figure in the legend refers to the finite-size scaling exponent. (c)~Distribution functions~(solid lines) of the largest accessibility index and the largest random walk centrality~(inverse accessibility index). They are compared with the Gumbel distribution function~(dotted lines) $P_G(x) = \exp\left[(x-\mu)/\beta - e^{-(x-\mu)/\beta}\right]/\beta$ with $\mu$ and $\beta$ being determined from the mean and the variance of the data.}
    \label{figA1}
\end{figure}

We investigate statistical properties of the random walk centrality or the accessibility index of sites in the 2D critical percolation cluster. Given a percolation cluster, we measure the accessibility indices $\alpha$ of all sites and construct a histogram of them normalized by the average value $\langle \alpha\rangle$. The probability  distribution function $P(\alpha/\langle \alpha\rangle)$ is then  obtained by taking the average of the histogram over the ensemble of percolation clusters. In Fig.~\ref{figA1}(a), we present and compare the distribution functions at three different values  of $L$. They overlap one another, which indicates that the accessibility distribution is characterized with only a single scale, namely the mean value $\langle\alpha\rangle$. The distribution function has asymmetric Gaussian tails.

We also study the system size dependence of the mean value $\langle \alpha\rangle$, the minimum value $\alpha_H$ of the hub, and the maximum value $\alpha_M$ of the marginal site. Their ensemble averaged values follow the power law $\alpha \sim L^{2.88(1)}$ with the same exponent close to the random walk exponent $\dw$. The accessibility index $\alpha_i$ has a meaning of the average MFPT to  $i$ from all the other sites. It is surprising that the average MFPT to  the most accessible site~(hub) and the least accessible site~(marginal site)  follow the finite-size scaling law with the same exponent.  This indicates that the average quantity alone is not a useful measure. The structural heterogeneity is not captured by the finite size scaling behavior of the average MFPT.

Figure~\ref{figA1}(c) presents the distribution functions of the maximum  accessibility index $\alpha_M/\langle\alpha\rangle$ and the maximum  random walk centrality $\langle\alpha\rangle / \alpha_H$ , normalized  with $\langle\alpha\rangle$.  The distribution functions are comparable with the Gumbel distribution which governs the extreme value statistics of Gaussian-distributed random variables.

\section{Distance Dependent MFPT}\label{App:theta_exp}
\begin{figure}
    \includegraphics*[width=\columnwidth]{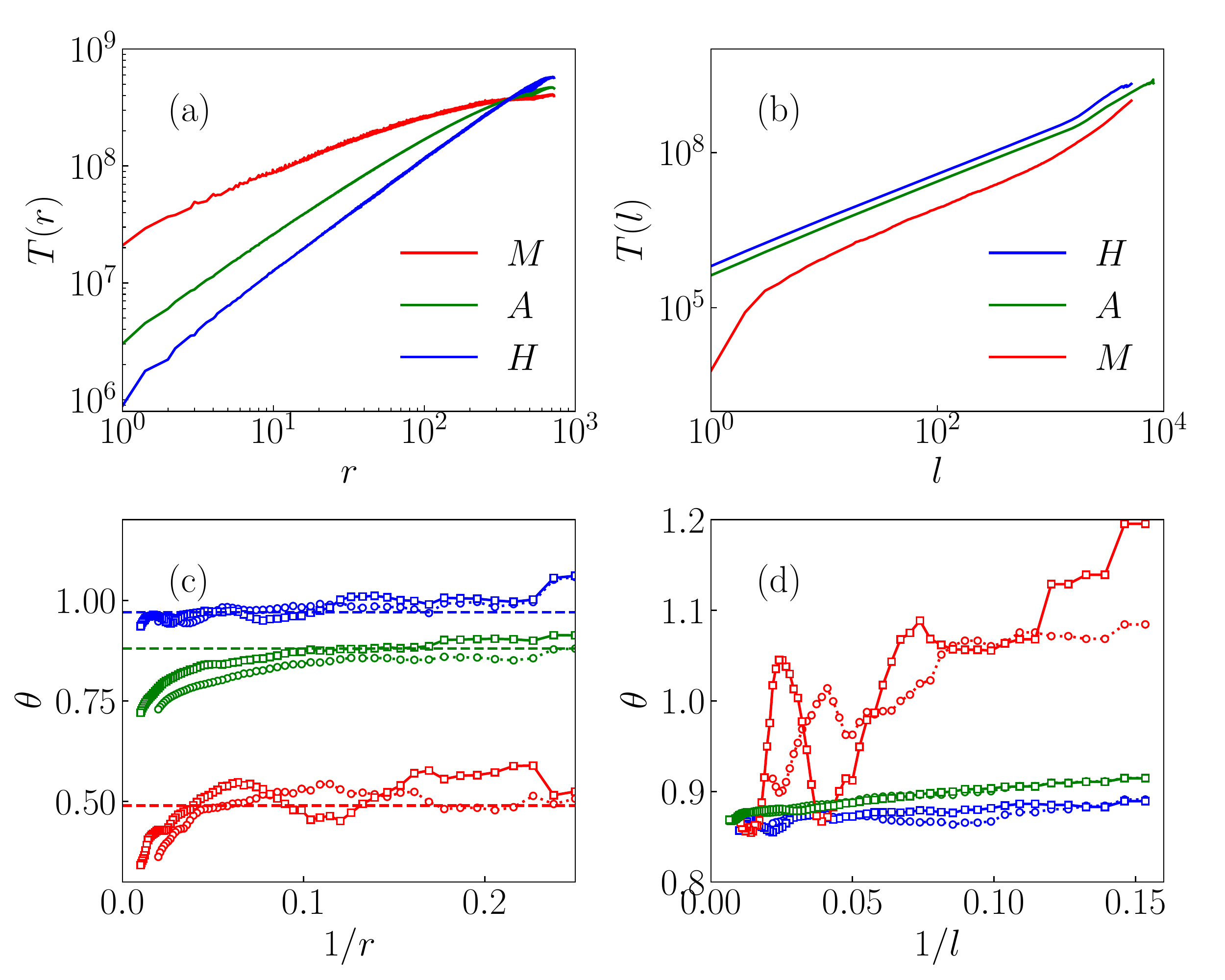}
    \caption{(a)~Euclidean distance~($r$) dependence and (b) chemical distance~($l$) dependence of the MFPTs from  the hub~(H), the marginal site~(M), and an arbitrary site selected at random~(A) at $L=1024$. (c) The effective exponents for the power law scaling  $T(r) \sim r^\theta$ at $L=512$~(circular symbols) and $L=1024$~(square symbols). The asymptotic values are marked with the dashed lines. (d) The effective exponents for the power law scaling $T(l) \sim l^{\theta_l}$ at $L=512$~(circular symbols) and $L=1024$~(square symbols).}
    \label{figA2}
\end{figure}
We have shown in the main text the MFPTs from the hub and the marginal site  follow a power law scaling
\begin{equation}
    T(r) \sim L^\Delta r^\theta
    \label{Tr_scaling}
\end{equation}
with distinct scaling exponents $(\Delta_h,\theta_h) \simeq (1.90, 0.97)$ for the hub and $(\Delta_m,\theta_m)\simeq(2.36, 0.49)$ for the marginal site.  We also measure $T_A(r)$, the average MFPT from an arbitrary source site selected {\em at random}~(see App.~\ref{App:randomSite}).  As shown in Fig.~\ref{figA2}(a), $T_A(r)$ also follows the scaling law of Eq.~\eqref{Tr_scaling}. The scaling exponent is obtained from an effective exponent analysis. The MFPTs within the range $r/\sqrt{2} < r_0 < \sqrt{2}r$ are fitted to yield the effective exponent $\theta(r)$.  Figure~\ref{figA2}(c) presents the effective exponents for $\theta_h$ for the hub, $\theta_h$ for the marginal site, and $\theta_r$ for a random site. The asymptotic scaling exponent is given by the limiting value in the  $r\to\infty$ and $L\to\infty$ limit. The effective exponents for $\theta_h$ and $\theta_m$ converge to the values obtained from the global fitting in the main text. The effective exponent for $\theta_r$ converges to $\theta_r \simeq 0.88$. The three distinct scaling exponents $\theta_h$, $\theta_m$, and $\theta_r$ signify the strong structural heterogeneity in the first passage processes.

The strong heterogeneity weakens when one adopts the chemical distance instead of the Euclidean distance as seen in Fig.~\ref{figA2}(b) and (d).  The chemical distance dependent scaling behavior will be discussed in App.~\ref{App:Tl}.

\section{MFPT from a random site}\label{App:randomSite}
\begin{figure}
\includegraphics*[width=\columnwidth]{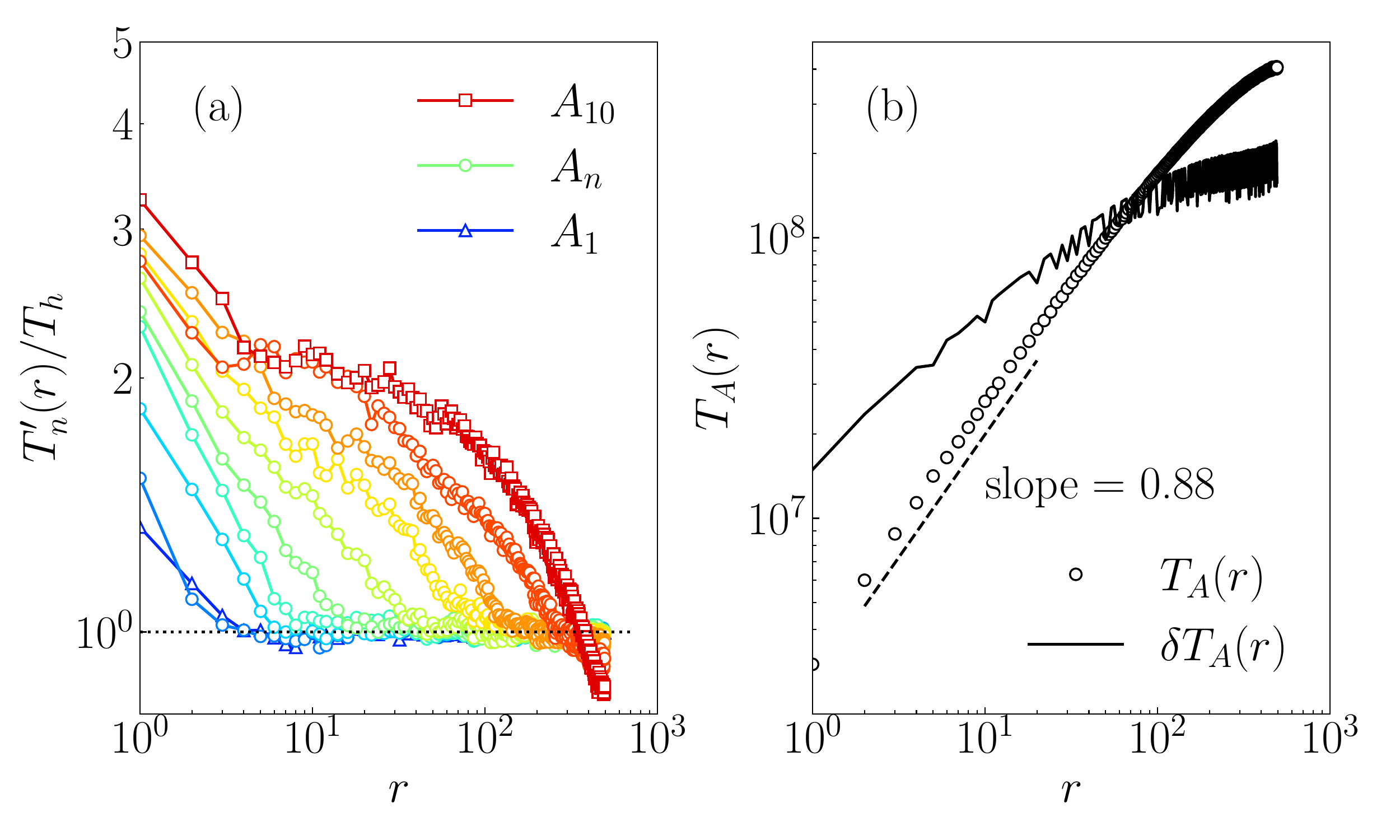}
\caption{(a) MFPT $T'_n(r)$ from an arbitrary site $A_n$  at a distance $R_n = 2^{n-1}$ from the hub normalized by $T_{h}(r)$. The horizon dotted line is a guide to the eye. (b) MFPT $T_A(r)$ from an arbitrary source site selected randomly.  If follows a power law scaling $T_A(r) \sim r^{\theta_r}$ with $\theta_r \simeq 0.88$. }
\label{figA3}
\end{figure}

We study the outbound MFPT from a set of sites $\{A_1, A_2, \cdots\}$ where $A_{n}$ is an arbitrary site selected  randomly among the sites at a distance $R_n = 2^{n-1}$ from the hub for 2D critical percolation clusters. We found that the MFPT $T'_n(r)$ from $A_n$ also exhibits the crossover scaling of the form~\eqref{crossover_scaling} of the main text.  In Fig.~\ref{figA3}(a), we plot the ratio of $T'_n(r)$ to $T_h(r)$, MFPT from the hub, to highlight the crossover. It clearly shows that $T'_n(r)/T_h(r)$ deviates from $1$ for $r\ll R_n$ and converges to $1$ for $r\gg R_n$.  Interestingly, the scaling exponent in the regime $r\ll R_n$ is given by $\theta_a \simeq 0.84(5) \neq \theta_m$.  It is close to $\theta_h$, but not the same. 

We also study outbound MFPT from a source site $A$ selected at {\em completely random} among all sites. In Fig.~\ref{figA3}(b), we compare the mean value $T_A(r)$ and the standard deviation $\delta T_A(r)$ of the MFPTs from $A$ to sites at a distance $r$. $T_A(r)$ corresponds to the MFPT averaged over all source-target pairs. It obeys a power law scaling $T_A(r)\sim r^{\theta_r}$ with a single exponent $\theta_r \simeq 0.88$, which is close to $d_w-d_f$. However, the standard deviation is even larger than the mean value for $r\ll L$. This strong non-self-averaging behavior is another indication of the heterogeneity.

\section{Direct and indirect explorations}\label{App:f_viaH}
We demonstrate a qualitative change in the ensemble of first-passage trajectories as the source-target distance increases. We focus on the 2D critical percolation clusters in $L^2$ square lattices. On a percolation cluster, we first determine the source site $s$ whose RWC is lowest among those at a distance $r_{h-s} = L/8$ from the hub, the maximum RWC site. We then choose randomly a target site $t$ among those at a distance $r_{s-t}$ from $s$. We generate $1024\times L$ random walk trajectories starting at $s$, and compute the fraction $f_{\rm viaH}$ of trajectories visiting the hub before arriving at $t$.

\begin{figure}
    \includegraphics[width=\columnwidth]{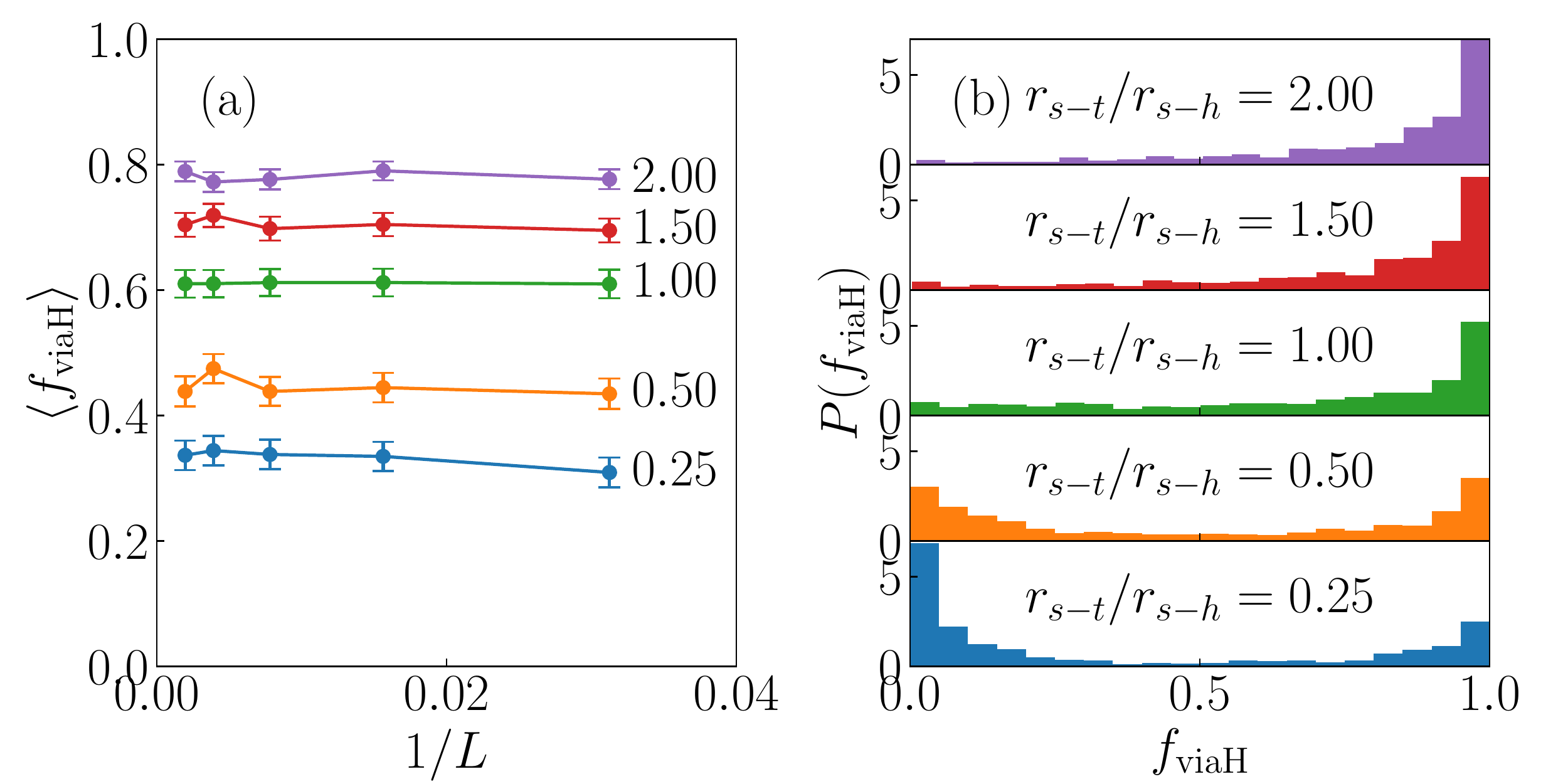}
    \caption{(a) Average fraction $\langle f_{\rm via H}\rangle$ of random walk trajectories starting from the local minimum RWC site $s$ and arriving at the target site $t$ via the hub. The distance between the hub and $s$ is $r_{s-h} = L/8$, and the relative distances between $s$ and $t$ are $r_{s-t}/r_{s-h} = 0.25, \cdots, 2.00$. (b) Distribution function of $f_{\rm via H}$ for $L=512$.}\label{figA4}
\end{figure}

Figure~\ref{figA4} (a) shows the fraction $\langle f_{\rm viaH}\rangle$ averaged over $1000$ percolation clusters. The average fraction seems to converge to a finite value in the large $L$ limit and increases gradually as $r_{s-t}$ increases. The probability of following indirect paths via the hub increases as $r_{s-t}$ increases. 

The distribution function shown in Fig.~\ref{figA4} (b) signifies a qualitative change. It is characterized by two peaks at $f_{\rm via H}=0$ and $1$, which reflect the dominance of direct and indirect paths via the hub, respectively. In the short distance regime~($r_{s-t} \lesssim r_{s-h}$), both peaks are evident. The double peak structure indicates strong sample-to-sample fluctuations in the ensemble of percolation clusters and the locations of source-target pairs. As $r_{s-t}/r_{s-h}$ increases, the peak at $f_{\rm via H}=0$ diminishes, leaving only a single peak at $f_{\rm viaH}=1$ in the long-distance regime~($r_{s-t} \gtrsim r_{s-h}$). The distribution function is clear evidence for the crossover from the direct exploration in the short distance regime to the indirect exploration via the hub in the long distance regime at the trajectory level.

\section{MFPT vs Chemical Distance}\label{App:Tl}

\begin{figure}
    \includegraphics[width=\columnwidth]{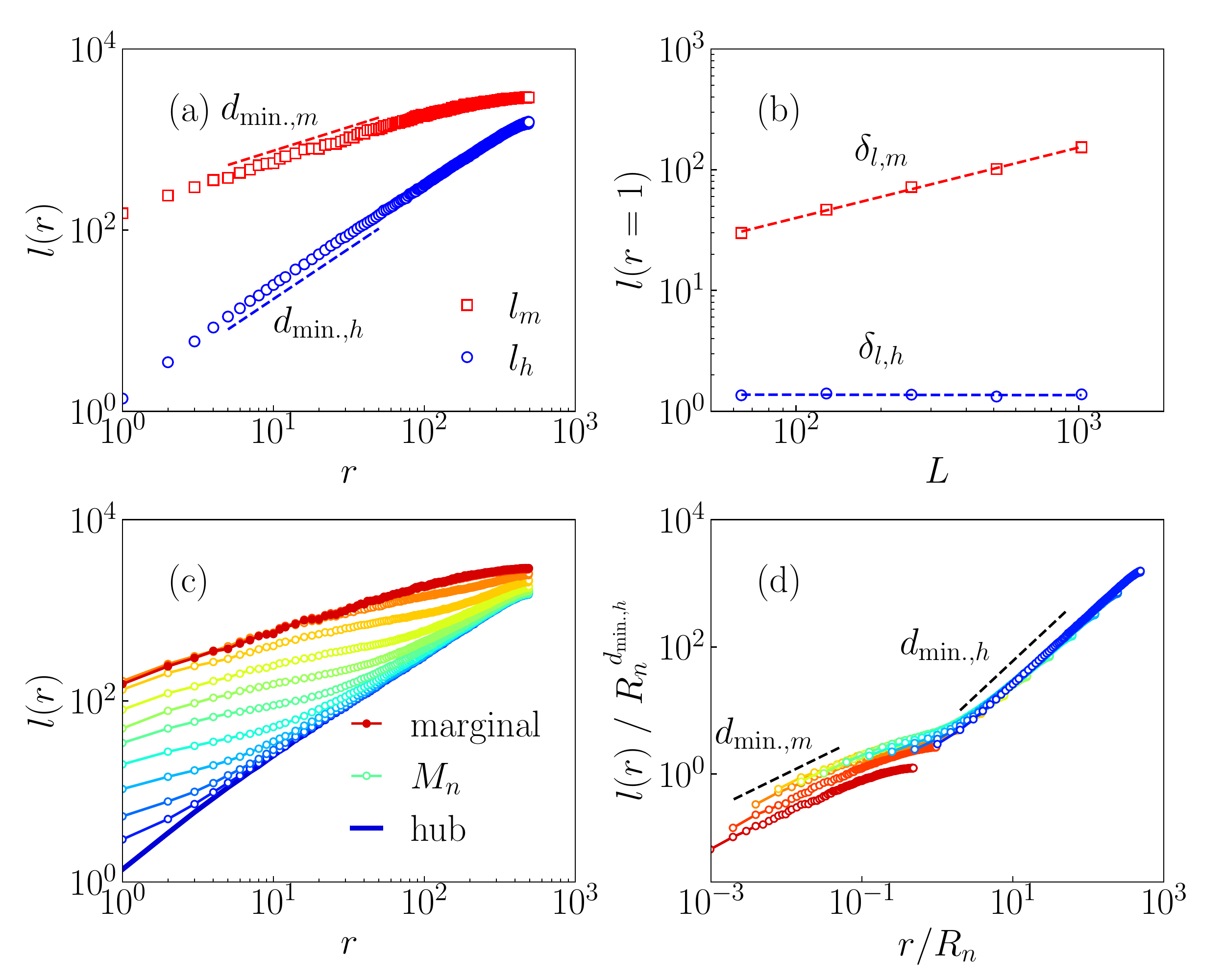}
    \caption{(a)~Chemical distance $l_{h/m}(r)$ from the hub/marginal site. (b)~System size $L$ dependence of $l_{h/m}(r=1)$. These plots confirm the power law scaling of Eq.~\eqref{lr_hm} with $\delta_{h}=0$, $\delta_{m}\simeq 0.52$, $d_{{\rm min.}, h} \simeq 1.11$, and $d_{{\rm min.},m}\simeq 0.58$. (c)~Chemical distance $l_n(r)$ from the local minimum RWC site $M_n$. (d)~Scaling plot of $l_n(r) / R_n^{d_{{\rm min.},h}}$ against $r/R_n$. $L=1024$ in (a), (c), and (d).}
    \label{figA5}
\end{figure}

Structural heterogeneity of the critical percolation cluster  is evidenced by the scaling behavior of the chemical distance $l(r)$ with respect to the Euclidean distance between two sites. The chemical distance between two sites is defined as the number of edges in the shortest path connecting them. It is known that the {\em average} chemical distance $l(r)$ between any pairs of sites at a distance $r$ scales as $l(r) \sim r^{d_{\rm min.}}$ with the chemical distance exponent $d_{\rm min.}$ (Sec.~6.6 of \cite{ben2000diffusion}),  which is for 2D critical percolation clusters $d_{\rm min.} \simeq 1.14$. We discriminate between the hub and the marginal site as starting site and find 
\begin{equation}
    l(r) \sim L^{\delta} r^{d_{\rm min.}}
    \label{lr_scaling}
\end{equation}
with {\em different} exponents $(\delta, d_{\rm min.}) =  (\delta_{h}, d_{{\rm min.},h}) \simeq  (0.0, 1.11)$ for the hub and $(\delta_{m}, d_{{\rm min.}, m}) \simeq (0.52, 0.58)$ for the marginal site~(see Figs.~\ref{figA5}(a) and (b)), satisfying $\delta_{h} + d_{{\rm min.}, h} \simeq \delta_{m} + d_{{\rm min.}, m}$. The chemical distance $l_n(r)$ from the local minimum RWC site $M_n$ shows again a crossover $l_n(r) = L^{\delta_h} R_n^{d_{{\rm min.}, h}} \mathcal{G}(l/ R_n)$ for $1 \ll R_n \ll L$ (see Figs.~\ref{figA5}(c) and (d)). 

\begin{figure}[t]
    \includegraphics[width=\columnwidth]{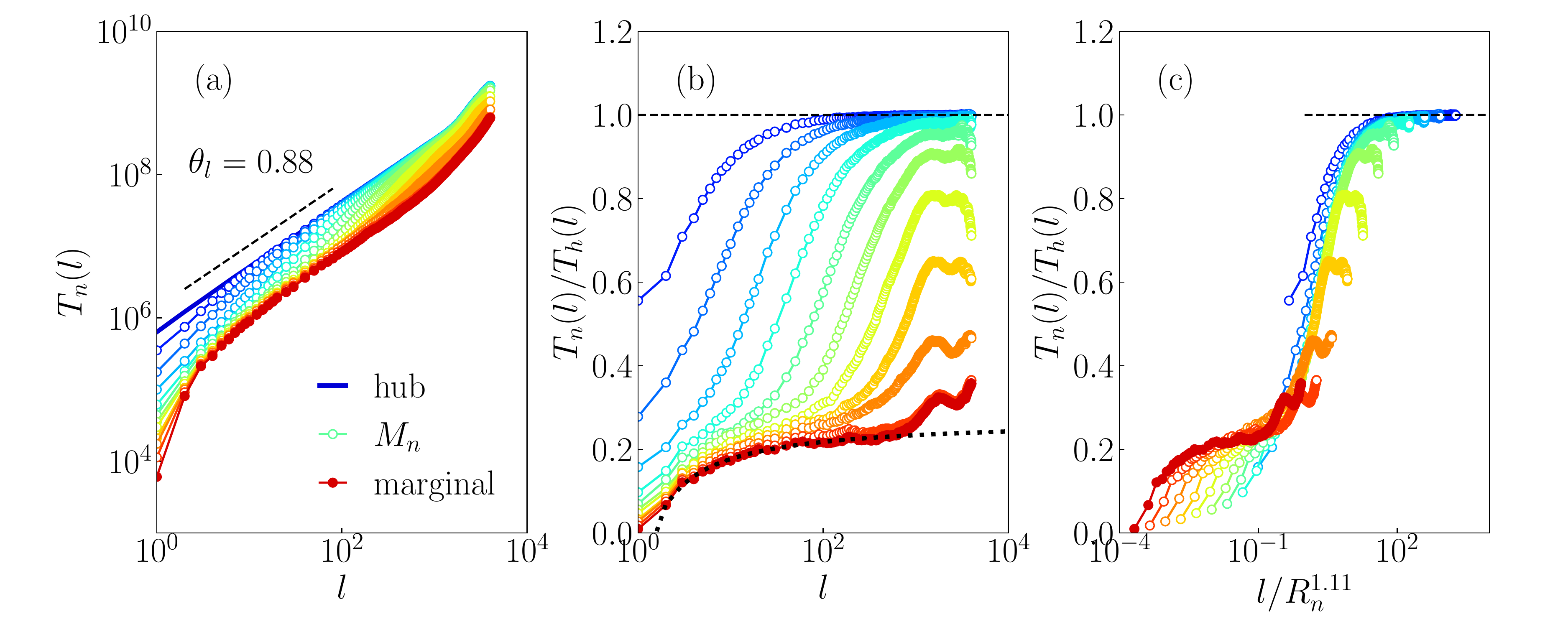}
    \caption{(a)~MFPT vs chemical distance from the hub and the local minimum RWC sites. (b)~Ratio of $T_n(l)$ to $T_h(l)$ for $n\geq 1$. For the marginal site, $T_{m}(l) / T_{h}(l)$ can be fitted to a function $0.27-0.26/\ln(1.70 l)$, which is drawn with a dotted line. (c) Scaling plot of $T_n(l)/T_h(l)$ against $l / R_n^{d_{\rm min.}}$.}
    \label{figA6}
\end{figure}

Given the similar crossover scaling of $T_n(r)$ and $l_n(r)$, one may anticipate a simple scaling, $T(l) \sim l^{\theta_l}$ with a unique exponent $\theta_l = \theta/d_{\rm min.}$, of the MFPT with respect to the chemical distance. Such a simple scaling behavior of the average MFPT was indeed reported in Ref.~\cite{benichou2008zero,condamin2008probing}. We have investigated the source site dependence of the MFPT function $T_n(l)$. Figure~\ref{figA6}(a) shows the $T_{h}(l)$ from the hub follows the power law scaling
\begin{equation}
    T_{h}(l) \sim l^{\theta_l}
    \label{T_h(l)}
\end{equation}
with $\theta_l \simeq 0.88$. This exponent satisfies the scaling relation $\theta_l = \theta_h / d_{{\rm min.},h}$.
The MFPT $T_n(l)$ from the local minimum RWC sites still displays a crossover, albeit weak. The crossover is more evident in Fig.~\ref{figA6}(b). $T_n(l)$ undergoes a crossover between two limiting behaviors, $T_{h}(l)$ of the hub and $T_{m}(l)$ of the marginal site, at a characteristic scale of the chemical distance $l_n \sim R_n^{d_{\rm min.}}$~(see Fig.~\ref{figA6}(c)).
The numerical data suggest that
\begin{equation}
    T_m(l) = T_h(l) / ( a - b/ \ln (cl))
\end{equation}
with $O(1)$ constants $a$, $b$, and $c$ (see Fig.~\ref{figA6}(b)). The logarithmic correction results in the slow convergence of the effective exponent in Fig.~\ref{figA2}(d).

    \section{Number of bridge bonds}\label{App:bridge}
\begin{figure}[t]
    \includegraphics*[width=\columnwidth]{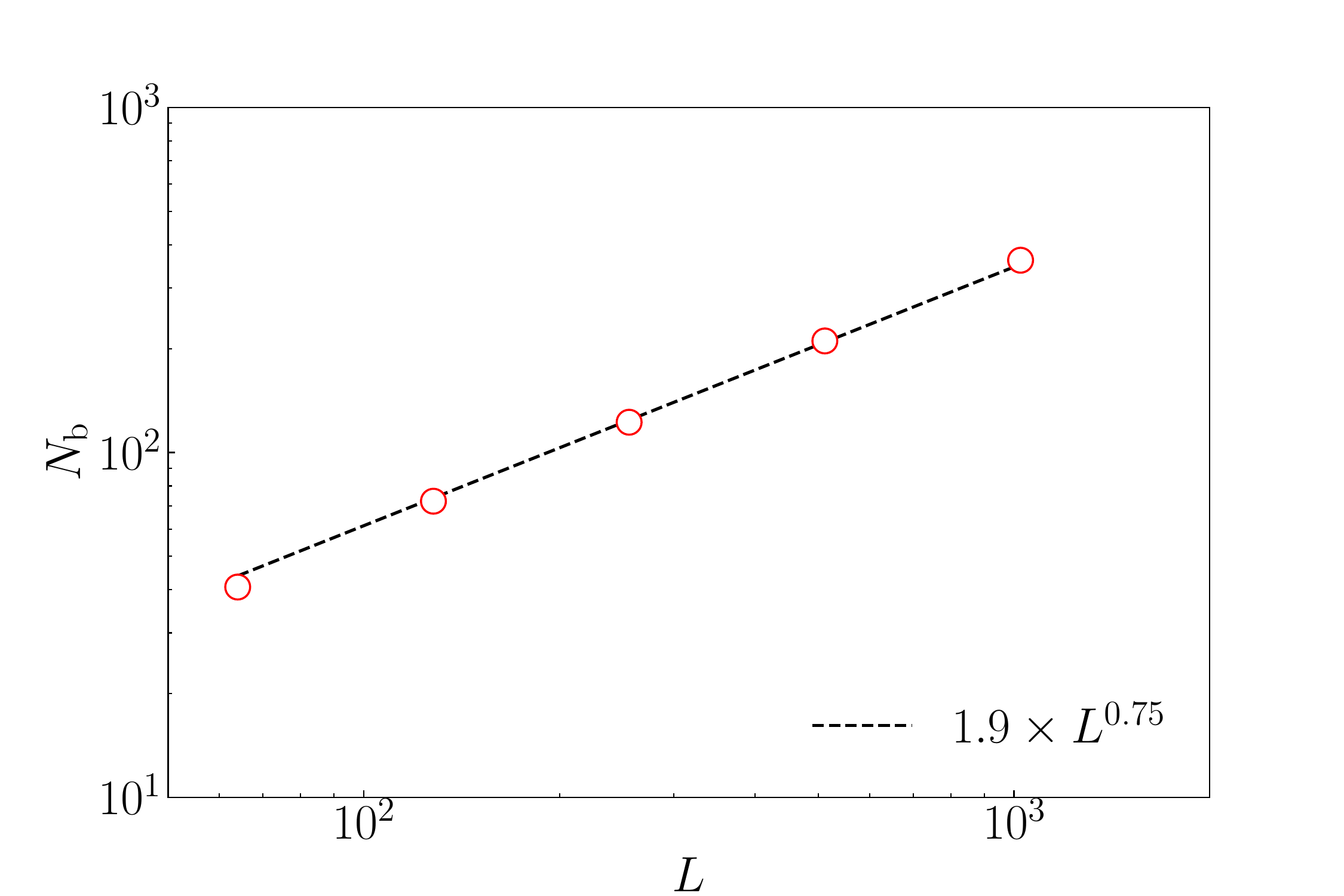}
    \caption{{The average number $N_{\rm b}$ of bridge bonds between the hub
    and the marginal site for on the 2D critical bond percolation clusters
on the $L\times L$ square lattices.}}\label{figA7}
\end{figure}

    We investigate numerically the scaling law for the number of bridge bonds between
    the hub~(highest RWC site) and the marginal site~(lowest RWC site) on
    the 2D critical bond percolation clusters. On a given realization of a
    critical percolation cluster on a $L\times L$ square lattice, we 
    identify the hub and the marginal site, and count the number of bridge
    bonds between the pair. The number of bridge bonds $N_{\rm b}$ averaged
    over $1000$ realizations are presented in Fig.~\ref{figA7}. It follows
    a power law $N_{\rm b} \sim L^{0.75}$ with respect to the system size
    $L$. We note that the scaling exponent coincides with the fractal
    dimension of the red bonds~\cite{Coniglio.1999}.

\section{Crossover scaling for 3D critical percolation clusters, 3D random trails, and the Sierpi\'nski gasket}\label{App:general}
\begin{figure}[t]
    \includegraphics*[width=\columnwidth]{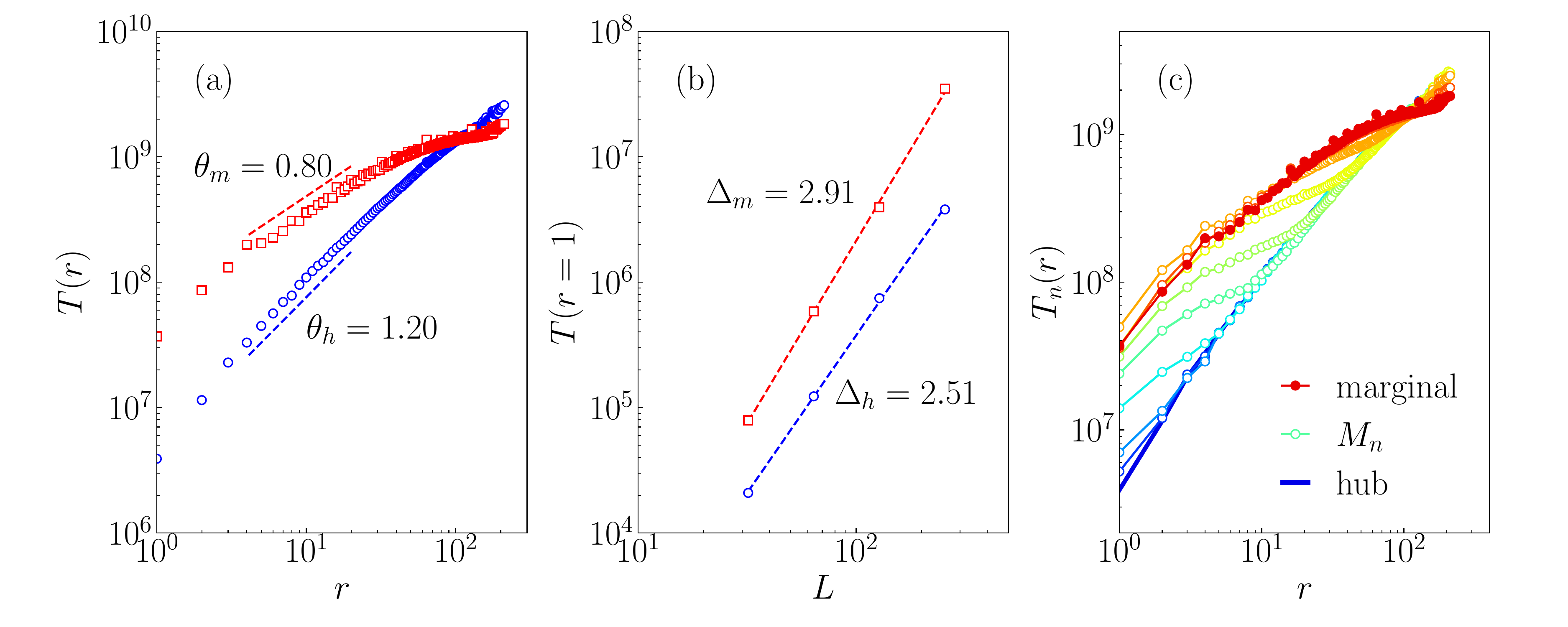}
    \caption{MFPT on the critical bond percolation cluster in 3D.  (a)~MFPT $T(r)$ from the hub and the marginal site for $L=256$. (b)~Size dependence of the MFPTs from the hub and the marginal site to the sites at a distance $r=1$. (c)~MFPTs from the hub, the marginal site, and the local minimum RWC sites for $L=256$. }
    \label{figA8}
\end{figure}

The crossover scaling is not limited to the 2D critical percolation clusters. To confirm the generality of the crossover scaling, we performed additional numerical studies.

First, we investigated the scaling law for the MFPT $T(r)$ on the 3D critical bond percolation clusters. It is known that the critical threshold is given by $p_c \simeq 0.248\ 812\ 6$ and the fractal dimension is given by $\df \simeq 2.523$~\cite{Lorenz.1998}. 
On a critical percolation cluster, we identified the hub~(highest RWC site),  $M_n$~(local minimum RWC site within a sphere of radius $R_n=2^{n-1}$ around the hub), and the marginal site~(lowest RWC site). The MFPTs from these sites to the other sites were measured as a function of the distance $r$. The ensemble averaged MFPTs from the hub and the marginal site satisfy the scaling law $T(r) \sim L^\Delta r^\theta$ with $(\Delta_h,\theta_h) \simeq (2.51, 1.21)$ for the hub and $(\Delta_m,\theta_m) \simeq (2.91, 0.79)$ for the marginal sites~(see Figs.~\ref{figA8}(a) and (b)).  The sums $\Delta_h+\theta_h$ and $\Delta_m+\theta_m$ are close to each other, and consistent with the random walk dimension $d_w \simeq 3.64\pm 0.05$~\cite{Havlin.2002}. Figure~\ref{figA8}(c) demonstrates that the MFPTs $T_n(r)$ from the local minimum RWC sites display the crossover scaling behavior.

Second, we examined the MFPT on a random walk trail in 3D, which is itself a random fractal with $d_f = 2$~\cite{Falconer.1985}. On an $L^3$ cubic lattice with periodic boundary conditions, we generated a random walk trail of $8L^2$ steps and explored the MFPT problem on top of it. 
Figure~\ref{fig4}~(b) of the main text illustrates the distance dependence of outbound MFPTs from the hub, the local minimum RWC sites, and the marginal site. The MFPT scales as $T(r) \sim L^{2.00} r^{1.25}$ for the hub and $T(r) \sim L^{1.80} r^{1.52}$ for the marginal site, and exhibits a crossover scaling for the intermediate sites.

\begin{figure}[t]
    \includegraphics[width=0.8\columnwidth]{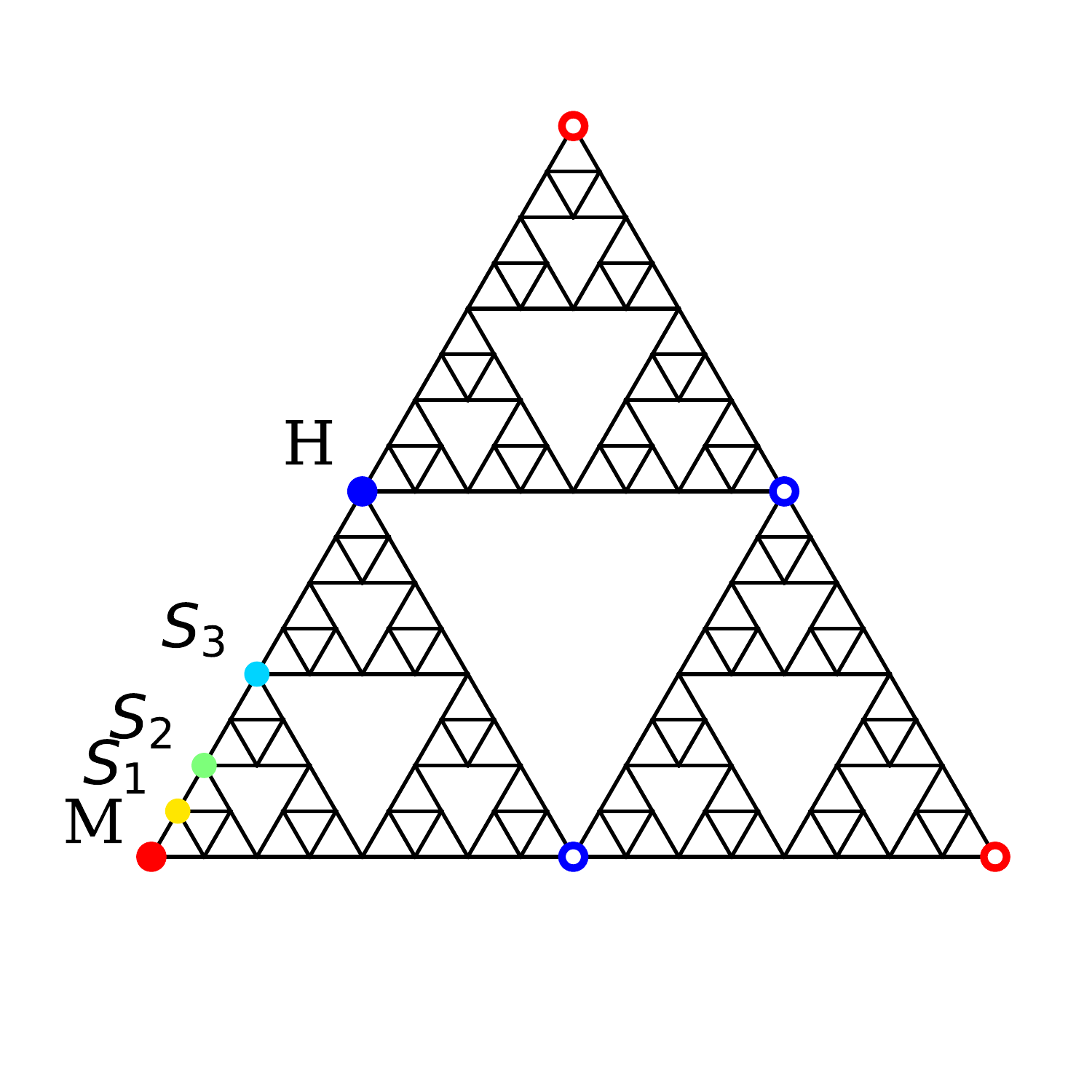}
    \caption{Sierpi\'nski gasket of the 5th generation. The three outermost sites are the marginal sites, and the three vertices of the largest empty triangle are the hubs. They are marked with red and blue symbols, respectively. The sites marked with filled symbols are the source sites for the study of MFPTs. An intermediate site $S_n$ is at a distance $2^{n-1}$ from the marginal site.}
    \label{figA9}
\end{figure}
Lastly, we studied the scaling of the MFPT on the Sierpi\'nski gasket, a deterministic fractal. Figure~\ref{figA9} depicts a Sierpi\'nski gasket of the 5th generation. As a deterministic fractal, it has a symmetric shape and a degenerate RWC distribution. To avoid ambiguity arising from this degeneracy, we considered the MFPTs from a hub, a marginal site, and intermediate sites $\{S_n\}$ on a straight line between them.
Figure~\ref{fig4}~(c) of the main text shows outbound MFPTs from the source sites on a Sierpi\'nski gasket of the 10th generation. 

The crossover scaling for the Sierpi\'nski gasket is weak in comparison with the other fractals. The MFPT obeys the power-law scaling $T(r) \sim A L^\Delta r^\theta$ with {\em site-independent} universal scaling exponents $\Delta = d_f$ and $\theta = d_w - d_f$ with $d_f = \ln3/\ln2$ and $d_w = \ln5 / \ln2$, where $L$ is a linear size. Interestingly, the amplitude $A$ displays a site-dependent crossover behavior between two values, $A_H$ for the hub and $A_M$ for the marginal site. The MFPT from $S_n$ scales with the amplitude $\sim A_H$ in a short distance regime and with the amplitude $\sim A_M$ in a long distance regime. The Sierpi\'nski gasket is a simple fractal characterized by a single fractal dimension $d_f$. It does not include any dangling ends, and there are not any bridge bonds between the hub and the marginal site. 
It {might explain} the reason why the scaling exponents are universal. Structural heterogeneity results in the weak crossover in the amplitude.

\end{document}